\documentclass[aps,prb,twocolumn,prabib,amsmath,amssymb,superscriptaddress]{revtex4-1}
\usepackage{graphics}
\usepackage{bm}
\usepackage{dcolumn}
\usepackage{natbib}
\usepackage{epstopdf}
\usepackage{float}
\usepackage{graphicx}
\usepackage{epsfig}
\usepackage[pdfstartview=FitH]{hyperref}
\usepackage{color}
\usepackage{appendix}
\usepackage{ulem}

\newcommand{\rr}{{\bf r}}

\begin{document}
\title{Fractional charge pumping of interacting bosons in one-dimensional superlattice}
\author{Tian-Sheng Zeng}
\affiliation{Department of Physics and Astronomy, California State University, Northridge, California 91330, USA}
\author{W. Zhu}
\affiliation{Department of Physics and Astronomy, California State University, Northridge, California 91330, USA}
\author{D. N. Sheng}
\affiliation{Department of Physics and Astronomy, California State University, Northridge, California 91330, USA}
\date{\today}
\begin{abstract}
Motivated by experimental realizations of integer quantized charge pumping in one-dimensional superlattices~[Nat. Phys. 12, 350 (2016); Nat. Phys. 12, 296 (2016)], we generalize and propose the adiabatic pumping of a fractionalized charge in interacting bosonic systems. This is achieved by dynamically sweeping the modulated potential in a class of one-dimensional interacting systems. As concrete examples, we show the charge pumping of interacting bosons at certain fractionally occupied fillings. We find that, for a given ground state, the charge pumping in a complete potential cycle is quantized to the fractional value related to the corresponding Chern number, characterized by the motion of the charge polarization per site. Moreover, the difference between charge polarizations of two ground states is quantized to an intrinsic constant revealing the fractional elementary charge of quasiparticle.
\end{abstract}
\maketitle
\section{introduction}
Quantized charge pumping via an adiabatically periodic perturbation is a hallmark of the two-dimensional quantum Hall effect~\cite{Klitzing1980},
which is connected to the Hall conductance by Laughlin's gedanken experiment~\cite{Laughlin1981,QNiu1985}.
Interestingly, it has been proven that the quantization of particle transport~\cite{Thouless1983}
is equivalent to the change of charge polarization~\cite{Vanderbilt1993,Ortiz1994,Vanderbilt2009,Wang2013,Zeng2015,Taddia2016}.
This intriguing relation stimulates recent progresses in cold atomic systems in one dimension,
as exemplified by experimental realization of the quantized charge pumping in both bosonic and fermionic atomic gases~\cite{Lohse2016,Nakajima2016}
following the theoretical proposal of quantized charge pumping in the superlattice~\cite{Wang2013a,Wei2015,Hatsugai2016} and the shaken bichromatic optical lattice~\cite{Mei2014}.

While most of previous studies were focused on the \textit{integer} quantized charge pumpings in one-dimensional cold-atom systems~\cite{Lohse2016,Nakajima2016}, it is interesting to ask that, is it possible to realize the fractional charge pumping in one-dimensional interacting system?
In two dimensions, fractional charge pumping usually occurs in the strongly-correlated systems with topological orders,
such as fractional quantum Hall (FQH) effects, where the quasiparticle excitations obey fractional anyonic statistics~\cite{Moore1991,Read1999,Read2000}.
Compared with the intense studies of topological orders in two dimensions, to our best knowledge,
the investigation of fractional charge pumping in one-dimensional interacting systems is still lacking and desired.  

As is well-known, topological ordered phases in two dimensions host topological degenerate ground states in the bulk,
which are indistinguishable by any local order parameter~\cite{Wen1990}. Topological degeneracy also implies the deconfined fractionalized quasiparticles emerging in the system.
Nevertheless, directly extending the definition of ``topological order'' from two dimensions to one dimension seems problematic,
since in one dimension interactions usually favor the translational symmetry breaking mechanism and generate crystalline orders.
However, as we will show below, there is no any obstacle to get one-dimensional phases with ``topological'' features similar to the two-dimensional topological ordered phases. Precisely, we will show the fractionalized charge pumping in these one-dimensional interacting systems, although the obtained new groundstates are locally distinguishable, which are not topological ordered phases~\cite{note1}. Interesting examples of phases with topological features in one dimension include FQH states in thin-torus limit, which can be smoothly connected to the intrinsic topological ordered phases in two dimensions~\cite{Seidel2005,Bergholtz2006}.

In the following discussion, we refer the one-dimensional phases as phases with topological features if the ground state carries non-trivial Berry phase and
realizes quantized charge pumping related to Chern number, despite the nearly degenerate ground states can be distinguishable by local density patterns, as these one-dimensional phases are descendant states of two-dimensional topological ordered phases.
Very recently, the possibility of such phases in one dimension has been studied in superlattices with periodically modulated potential~\cite{Lang2012,Kraus2012,Zhu2013,Ganeshan2013,Grusdt2013}.
In such quasiperiodic lattices, the topological property of the system can be understood
in terms of a nonzero integer Chern number which is robust even under the inclusion of disorders~\cite{Lang2012,Kraus2012},
with the appearance of edge states equivalent to the edge states of a two-dimensional integer quantum Hall system. In the presence of interactions (c.f.~the dipolar interaction), it was noticed that there exists $m$-fold quasi-degenerate ground states with nontrivial quantized Chern number and commensurate crystalline orders at fractional filling factors $\nu=1/m$ in these quasiperiodic lattices~\cite{Guo2012,Xu2013,Budich2013,Li2015,Hu2015,Regnault2012}, which is due to one-dimensional crystalline nature of two-dimensional Abelian FQH states in thin torus limit. The topological properties of these states are characterized by (\textrm{i}) fractionally quantized Chern invariants, (\textrm{ii}) degenerate ground state manifolds under the adiabatic insertion of flux quanta, and (\textrm{iii}) the quasihole exclusion statistics.

In this work, we study the fractional charge pumping of interacting bosons in a fractionally occupied
one-dimensional superlattice through exact diagonalization (ED) and density-matrix renormalization group (DMRG) methods. We show that one dimensional quasi-degenerate ground states with a nontrivial manybody Chern number and fractionally charged quasihole excitations, emerge at given fillings $\nu=k/2$ of topological band with $(k+1)$-body interactions, which are counterparts to two-dimensional non-Abelian FQH states in thin torus limit, but trivially distinguishable by local density patterns. Their nontrivial quantized features, with certain interesting features similar to the Hall conductance of two-dimensional counterparts, can be revealed by fractional quantization of charge pumping, which is related to the motion of charge polarization.

\section{The Model Hamiltonian}

In cold atomic systems, one-dimensional superlattice is formed by superimposing two lattices with different wavelengths~\cite{Lohse2016,Fallani2007}, whose tight binding Hamiltonian can be written as
\begin{align}
  H_0=\sum_{\langle ij\rangle}-t(b_{i}^{\dag}b_{j}+h.c.)+\sum_{j}\mu\cos(2\pi\phi j+\theta) n_{j}. \label{eq:ham0}
\end{align}
Here, $n_{j}=b_{j}^{\dag}b_{j}$, $\mu=-2t$ is the onsite potential modulation amplitude, and $\phi=1/q,q=3$ is the commensurate periodic factor. The hopping $t$ can be tunable through laser-assisted tunnelling~\cite{Gerbier2010,Aidelsburger2011}. The energy spectrum of $H_0$ with varying $\theta\in [0,2\pi]$ is identical to that of a two-dimensional Hofstadter model~\cite{Kraus2012,Lang2012}, and hosts nontrivial topological invariant $|C|=1$ in the lowest Hofstadter subband~\cite{Hofstadter}. With bosonic polar molecules $^{87}\text{Rb}^{133}\text{Cs}$~\cite{Takekoshi2014} and $^{23}\text{Na}^{87}\text{Rb}$~\cite{Guo2016} loaded into this setup, the effective interaction potential between site $i$ and site $j$ in a circularly polarized microwave field, reduces to the form including  two-body interactions $D(i-j)n_i n_j$, three-body interactions $V(i-j)n_{i}^{2}n_j$ and higher-order four-body interactions $W(i-j)(n_{i}^{2}n_{j}^{2}+n_{i}^{3}n_j)$ in the perturbation expansion of dipolar interaction, as
suggested in Ref.~\cite{Zoller2007a}. With the onsite contact interactions~\cite{Lewenstein2012}, these bosons experience the extended $n$-body ($n\leq4$) Hubbard interactions $\sum_{j,n}U_n \prod_{k=0}^{n-1}(n_j-k)+\sum_{i\neq j}D(i-j)n_i n_j+H_3+H_4$, with three-body terms
\begin{align}
  H_3=&\sum_{i\neq j}V(i-j)\Big[n_j(n_j-1)n_i+i\leftrightarrow j\Big],
\end{align}
and four-body terms
\begin{align}
  H_4=&\sum_{i\neq j}W(i-j)\Big[\prod_{k=0}^{2}(n_j-k)n_i+i\leftrightarrow j\nonumber\\
      &+n_i(n_i-1)n_j(n_j-1)\Big].
\end{align}
While the two-body parts $D(i-j)$ can be tuned down to a small value by manipulating microwave fields~\cite{Zoller2007a,Zoller2007b}, we focus our interest on the $n$-body ($n\geq3$) long range interactions. Due to the rapidly decaying of dipolar potential, we truncate the interaction terms by cutting off the tails $|i-j|>3$. Thus, at fixed filling $\nu=N\times q/L$ of the lowest Hofstadter subband, where $N$ is the particle number and $L$ the chain length, as the detailed parameters $V_i,W_i$ are rather complicated (see also Appendix D), we choose typical interaction parameters $U_3=\infty,V(i)=V_i$ and $U_4=0,W(i)=0$ at $\nu=1$ and $U_4=\infty,W(i)=W_i$ and $U_3=V(i)=0$ at $\nu=3/2$ to simplify the calculation, and also to claim the physical importance of long range interaction $V_i,W_i$ for the existence of non-trivial ground states we consider here. In ED calculations, the largest accessible cluster is 12 particles and the dimension of Hilbert space is of the order of $10^{8}$. With the translational symmetry, the energy states are labeled by the total momentum $(K,\theta)$ in the magnetic
Brillouin zone. For DMRG, we keep the number of states more than $800$, and the truncation error is less than $10^{-8}$ to ensure accurate results.

\section{ground states}

\begin{figure}[b]
  \includegraphics[height=1.55in,width=3.4in]{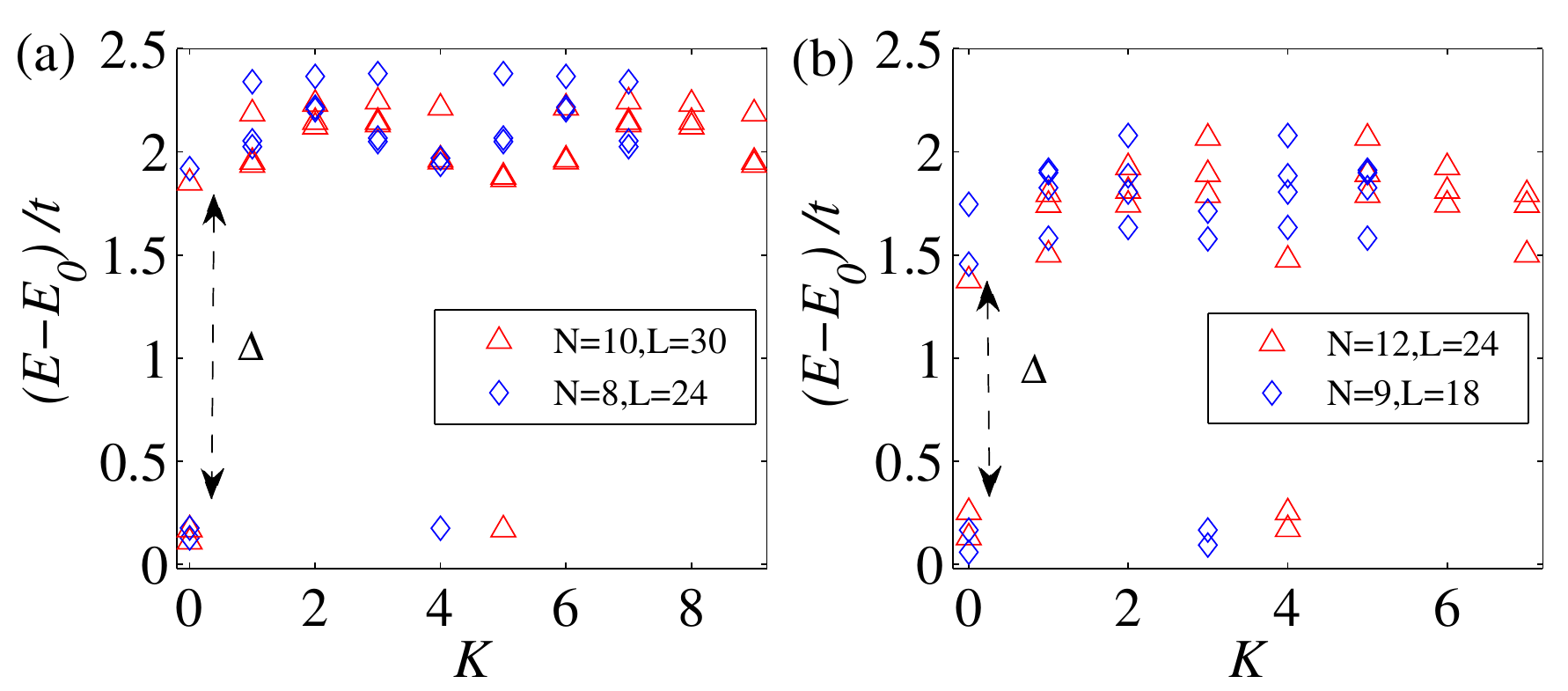}
  \caption{\label{ground}(Color online) Low energy spectrum for (a) three-body hard-core bosons $H_0+H_3$ at $N=8,L=24,U_3=\infty,V_1=V_2=V_3=10t$; (b) four-body hard-core bosons $H_0+H_4$ at $N=9,L=18,U_4=\infty,W_1=W_2=W_3=10t$. $\Delta$ marks the energy gap between quasi-degenerate ground state manifolds and excited levels. Here $\theta=0$.}
\end{figure}

In two dimensions, interacting bosons are predicted to form incompressible liquids
at filling factor $\nu=k/2$ ($k\in Z_+$) with $(k+1)$-body interactions~\cite{Read1999,Cooper2001}. Most notable examples include Moore-Read (MR) states at $\nu=1$ ($k=2$) and Read-Rezayi (RR) states at $\nu=3/2$ ($k=3$)~\cite{Rezayi2005,Zhu2014}.
Remarkably, the corresponding topological nature is captured by the analysis of ``root configuration''
or generalized Pauli principle in ``thin-torus limit'': No more than $k$ bosons in two consecutive orbitals~\cite{Bernevig2008,Emil2008,Papic2014}. By squeezing the geometry into the thin-torus limit, these quantum Hall states are adiabatically connected to a charge density wave characterized by the root configuration~\cite{Seidel2005,Bergholtz2006,Ardonne2008}.

For one-dimensional systems we studied, we demonstrate their ground states are quasi-degenerate charge density wave and host nontrivial Berry phase, which is the counterpart of non-Abelian FQH states in thin torus limit. We first obtain the low-energy spectrum of the $H_0+H_3$ and $H_0+H_4$ at filling numbers $\nu=1$ and $\nu=3/2$, respectively. As shown in Figs.~\ref{ground}(a-b), we find strong numerical evidence of threefold quasi-degenerate ground states (two lie in $K=0$ sector while the other in $K=\pi$ sector) for even particle numbers at $\nu=1$, in the condition of strong three-body interactions $V_i\gg t$. Similarly, fourfold quasi-degenerate ground states (two lie in $K=0$ sector while the other two in $K=\pi$ sector) is found at filling $\nu=3/2$ in the interaction dominate regime.

\begin{figure}[t]
  \includegraphics[height=2.8in,width=3.4in]{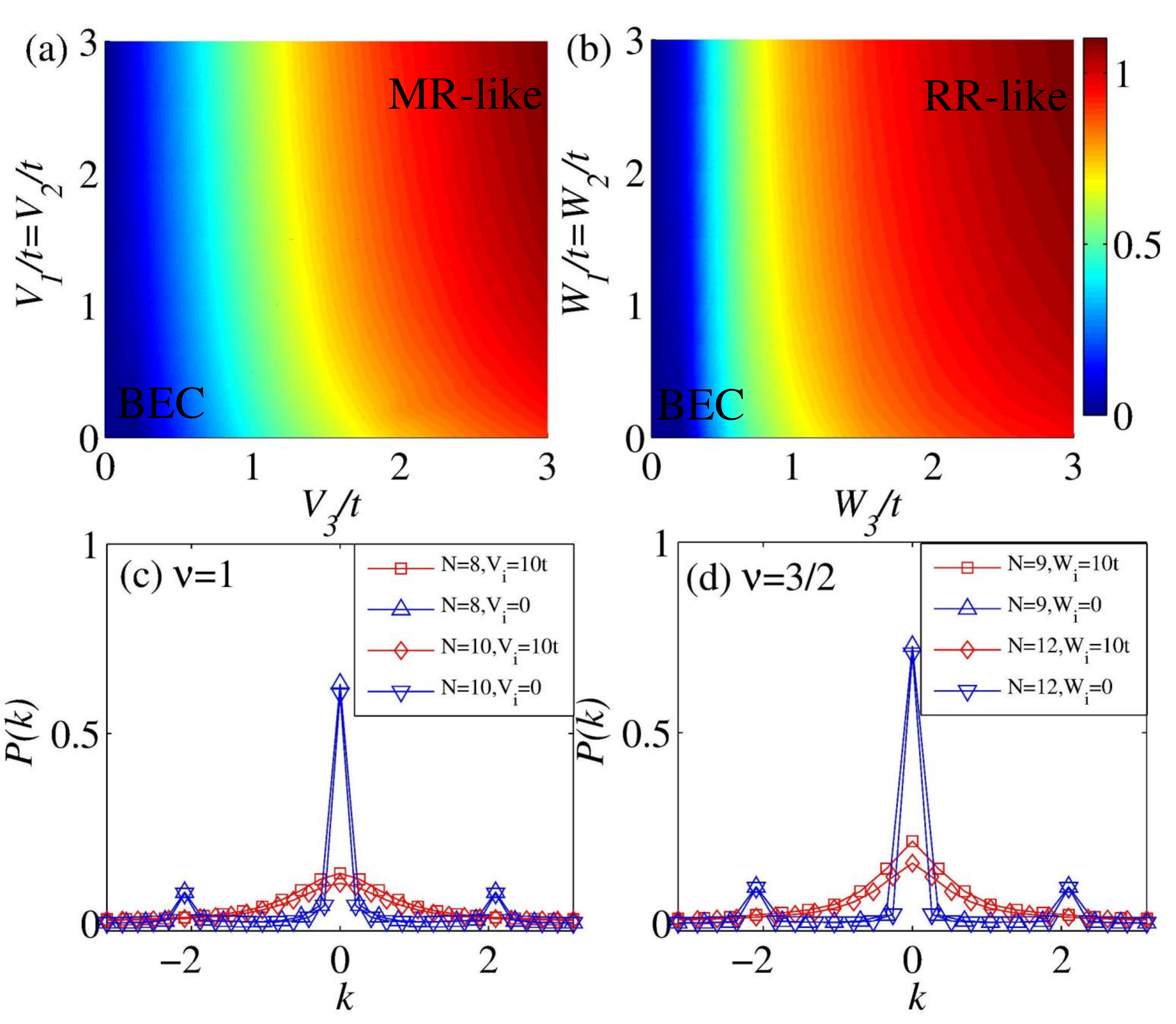}
  \caption{\label{energy}(Color online) Contour plot of energy gap $\Delta$ versus interaction at $\theta=0$ for (a) three-body hard-core bosons $H_0+H_3$ at $\nu=1,N=8,L=24,U_3=\infty$; (b) four-body hard-core bosons $H_0+H_4$ at $\nu=3/2,N=9,L=18,U_4=\infty$. Diffraction pattern $P(k)$ of the lowest ground state for (c) three-body hardcore bosons at $\nu=1$ and (d) four-body hardcore bosons at $\nu=3/2$.}
\end{figure}

Moreover, by calculating the Berry curvatures using mesh points $m\times m$ with $m\geq9$, it is found that  the sum of the many-body Chern number $C_{\alpha}$ of three (four) gapped ground states is equal to a constant number $C=\sum_{\alpha}C_\alpha=3$ ($C=6$) at $\nu=1$ ($\nu=3/2$). Based on the above results, we interpret these one-dimensional ground states as MR-like states at $\nu=1$ (RR-like states at $\nu=3/2$) throughout our discussions here, since obviously all topological features in such an one-dimension system are inherited from two-dimensional topological ordered $\nu=1$ MR ($\nu=3/2$ RR) states (see also the Appendix~\ref{app1} for quasihole excitations). In addition, these ground states also display commensurate Bragg peaks in the density structure factor, which agree with their corresponding root configurations. As we will show below, despite these ground states can be distinguishable by local density patterns, the non-trivial Chern number promises the quantized charge pumping effect.

To explore the stability of these phases, we define the the energy gap $\Delta$ as the difference between the highest energy of the quasi-degenerate ground states and the first excited energy. As shown in Figs.~\ref{energy}(a-b), the energy gap opens for strong interactions $V_3\gg t$ ($W_3\gg t$), while it closes for very weak interactions $V_3\ll t$ ($W_3\ll t$), which is possibly a metallic phase. To verify it, we compute the off-diagonal long range order $\rho_{ij}=\langle b_{i}^{\dag}b_{j}\rangle$~\cite{Mueller2006} of the lowest ground state. By diagonalizing the $L\times L$-matrix $\rho_{ij}$, we obtain reduced single particle eigenstates $\rho|\phi_{\alpha}\rangle=\rho_{\alpha}|\phi_{\alpha}\rangle$ where $|\phi_{\alpha}\rangle$ ($\alpha=1,\ldots,L$) are the
effective orbitals as eigenvectors for $\rho$ and $\rho_{\alpha}$ ($\rho_1\geq\ldots\geq \rho_{L}$) are interpreted as occupations. We find that: (\textrm{i}) the occupations $\rho_{\alpha}\simeq N$ for $\alpha=1$, while $\rho_{\alpha}\ll1$ for $\alpha>1$ in the weakly interacting regime; (\textrm{ii}) $\rho_{\alpha}\simeq \nu$ for $\alpha\leq N/\nu$, while $\rho_{\alpha}\ll1$ for $\alpha>N/\nu$ in
the strongly interacting regime, demonstrating the development of strong correlation. Figs.~\ref{energy}(c-d) show the diffraction pattern
\begin{align}
  P(k)=\frac{1}{N L}\sum_{j,j'}\rho_{jj'}e^{i k\cdot(j'-j)},
\end{align}
whose peak position signaling the condensation momentum. For $V_3\ll t$ ($W_3\ll t$), $P(k)$ has a sharp peak at momenta $k=0$, indicating a Bose-Einstein condensate (BEC), while $P(k=0)$ for $V_3\gg t$ ($W_3\gg t$) tends to diminish by increasing the system sizes.

\section{Charge Pumping}

\begin{figure}[t]
  \includegraphics[height=3.0in,width=3.4in]{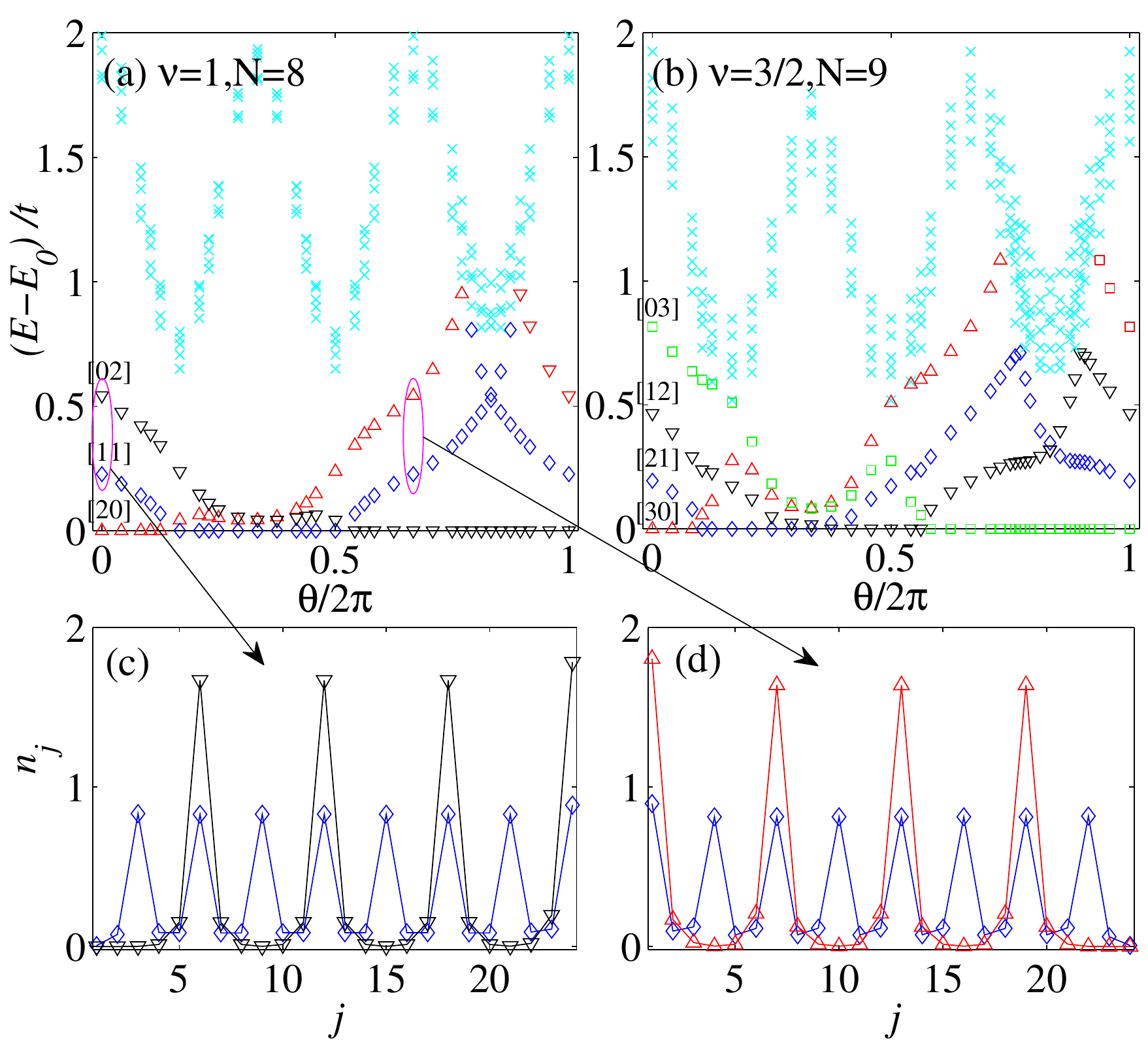}
  \caption{\label{edge}(Color online) Energy spectra vs the variation of $\theta$ under open boundary conditions for (a) three-body hard-core bosons $H_0+H_3$ at $N=8,L=24,U_3=\infty,V_1=V_2=V_3=10t$; (b) four-body hard-core bosons $H_0+H_4$ at $N=9,L=18,U_4=\infty,W_1=W_2=W_3=10t$. The topological sectors of the low energy states at $\theta=0$, identified by the density pattern (root configuration), are labeled in the parentheses. We only show ten lowest energy levels at each $\theta$. (c-d) The density distributions for the boundary states at $\theta=0$ and $\theta=4\pi/3$, respectively.}
\end{figure}

The nontrivial topological properties in the bulk are closely related to the edge physics.
For MR-like and RR-like states, in Figs.~\ref{edge}(a-b), we plot the low energy spectrum under open boundary by dropping off the boundary hopping term $-t(b_{L}^{\dag}b_{1}+h.c.)$. In contrast to the case of periodic boundary condition, the degenerate manifold of the ground states is lifted, and there are low energy states filling in the gap. In one-dimensional limit, according to the analysis of Refs.~\cite{Seidel2005,Bergholtz2006,Ardonne2008}, the root pattern coincides with the charge-density-wave (CDW) pattern. Thus we can distinguish the different sectors of these ground states by measuring their density patterns. For example, in Figs.~\ref{edge}(a-b), the ground states are labeled by $[20]$, $[02]$ and $[11]$ for $\nu=1$, and $[30]$, $[03]$, $[21]$ and $[12]$ for $\nu=3/2$. These states which reside within the gap, are adiabatically connected to the excited levels as $\theta$ is tuned. They have edge excitations localized either on the left or on the right boundary of the chain, as indicated in Figs.~\ref{edge}(c-d).
Moreover, different from Abelian FQH states with only one chiral edge branch (the chirality is defined by the sign of current $J=\nabla_{\theta}E(\theta)$),
we observe that there are more than one edge branches for these nontrivial states, as those of two-dimensional counterpart states~\cite{Liu2012}.

\begin{figure}[t]
  \includegraphics[height=1.6in,width=3.4in]{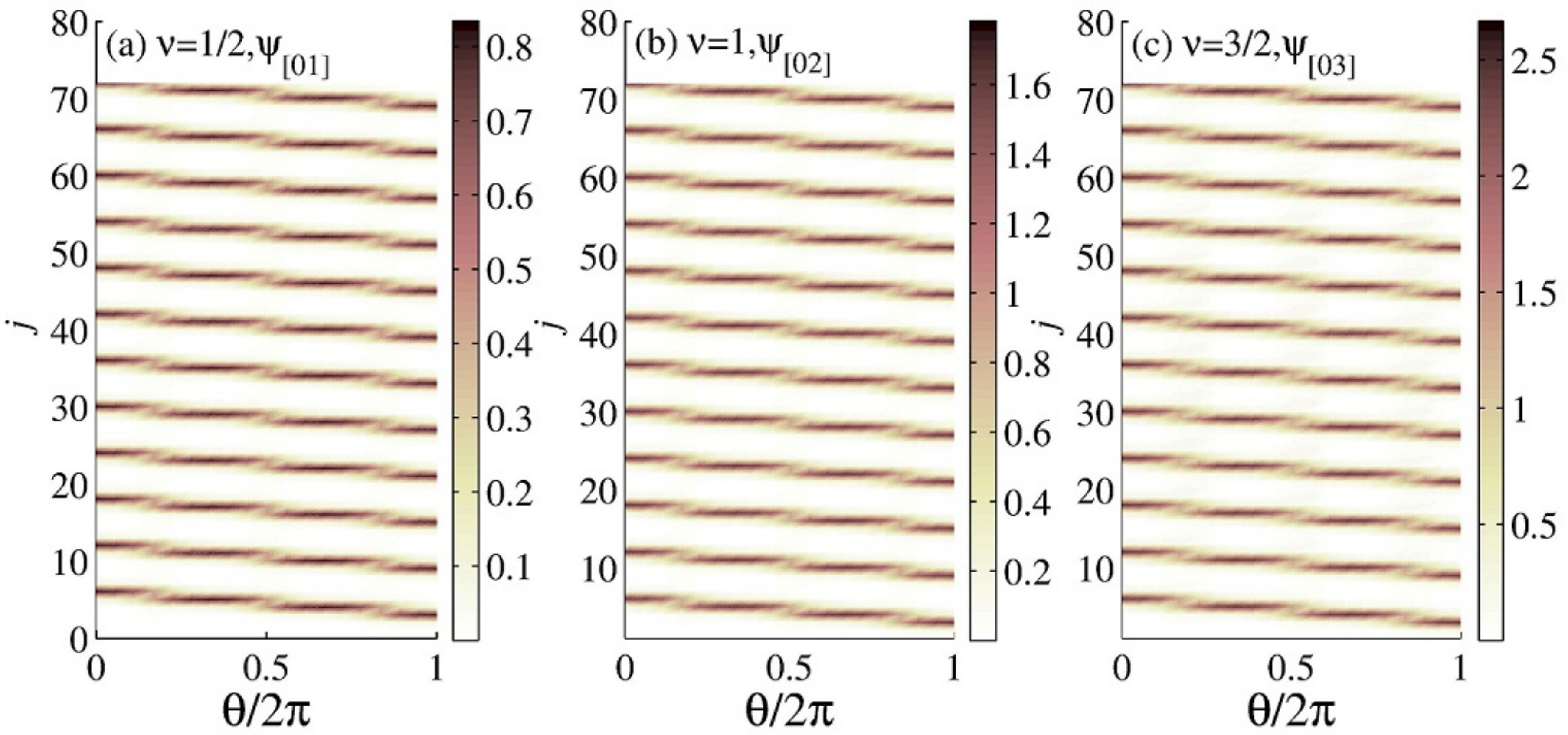}
  \caption{\label{center}(Color online) Density distributions $\langle\psi_{\alpha}(\theta)|n_{j}|\psi_{\alpha}(\theta)\rangle$ of a specific ground state along the evolution path $\theta$ for (a) two-body hard-core bosons $H_0+\sum_{i,j}D(i-j)n_i n_j$ at $\nu=1/2,N=25,L=150,U_2=\infty,D(1)=D(2)=D(3)=2t$; (b) three-body hard-core bosons $H_0+H_3$ at $\nu=1,N=24,L=72,U_3=\infty,V_1=V_2=V_3=10t$; (c) four-body hard-core bosons $H_0+H_4$ at $\nu=3/2,N=36,L=72,U_4=\infty,W_1=W_2=W_3=10t$. The cloud shifts globally, and the center-of-mass position is quantized (see Fig.~\ref{pump}).}
\end{figure}

In order to visualize the topological nature of these ground states, we ramp linearly $\theta$ from $0$ to $2\pi$
and inspect the dynamical charge pumping process, which can be obtained by a recently developed adiabatic DMRG technique~\cite{He2014,Zhu2015}. At typical $\theta$, we show the evolution of the density distributions vs $\theta$ of one selected ground state at different fillings, as illustrated in Figs.~\ref{center}(a-c). We see that, despite the CDW pattern persists from $\theta=0$ to $\theta=2\pi$, the density configuration shifts globally. When the phase $\theta$ equals to the integer multiples of $2\pi/q$, namely $\theta=2\pi p/q$, the ground state recovers the configuration at $\theta=0$, with the density configuration shifting $p$ lattice sites towards the $j=1$ end, which can be attributed to the $2\pi/q$-periodic Berry curvature~\cite{Marra2015}.

\begin{figure}[t]
  \includegraphics[height=1.5in,width=3.2in]{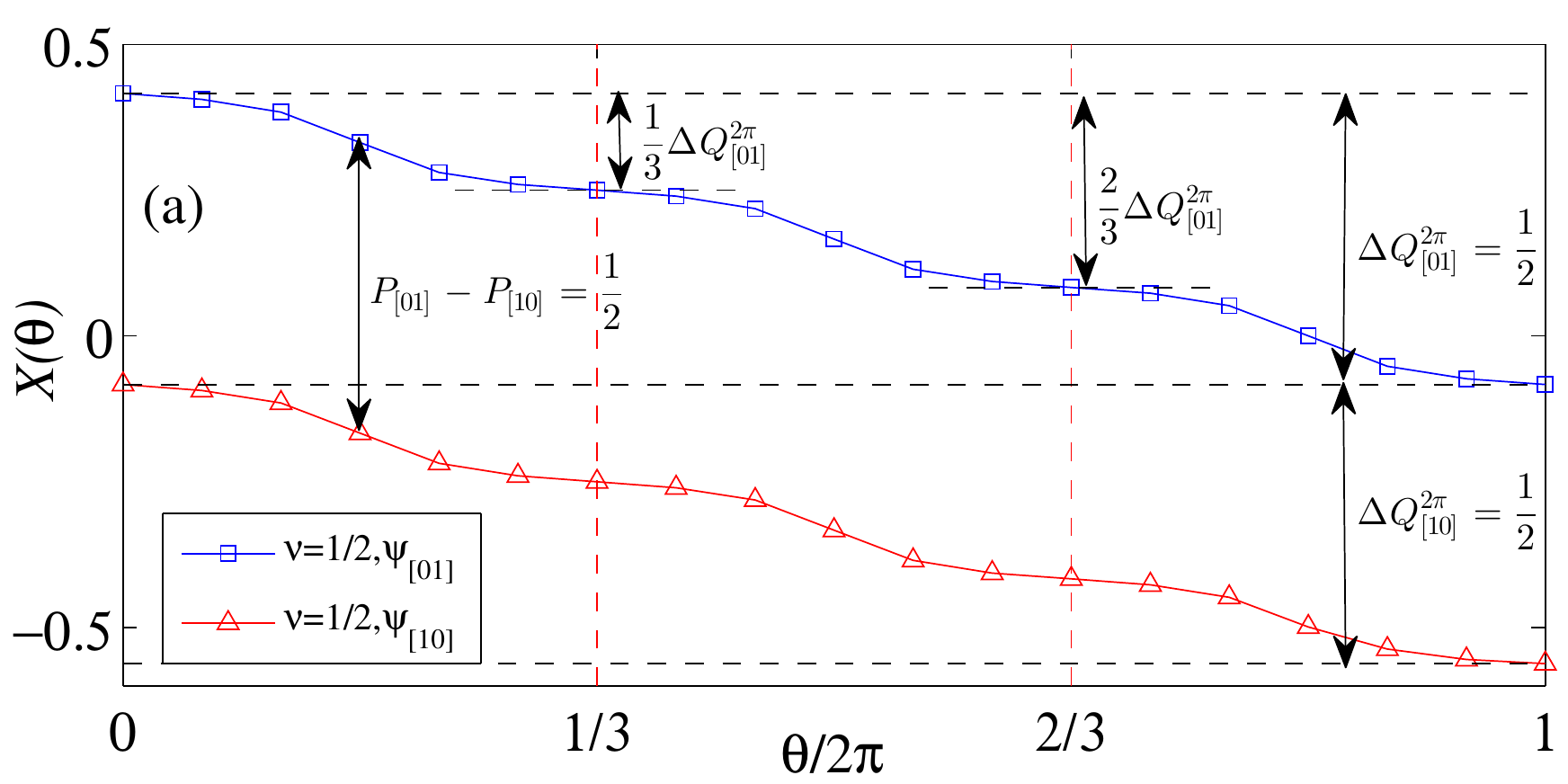}
  \includegraphics[height=1.5in,width=3.2in]{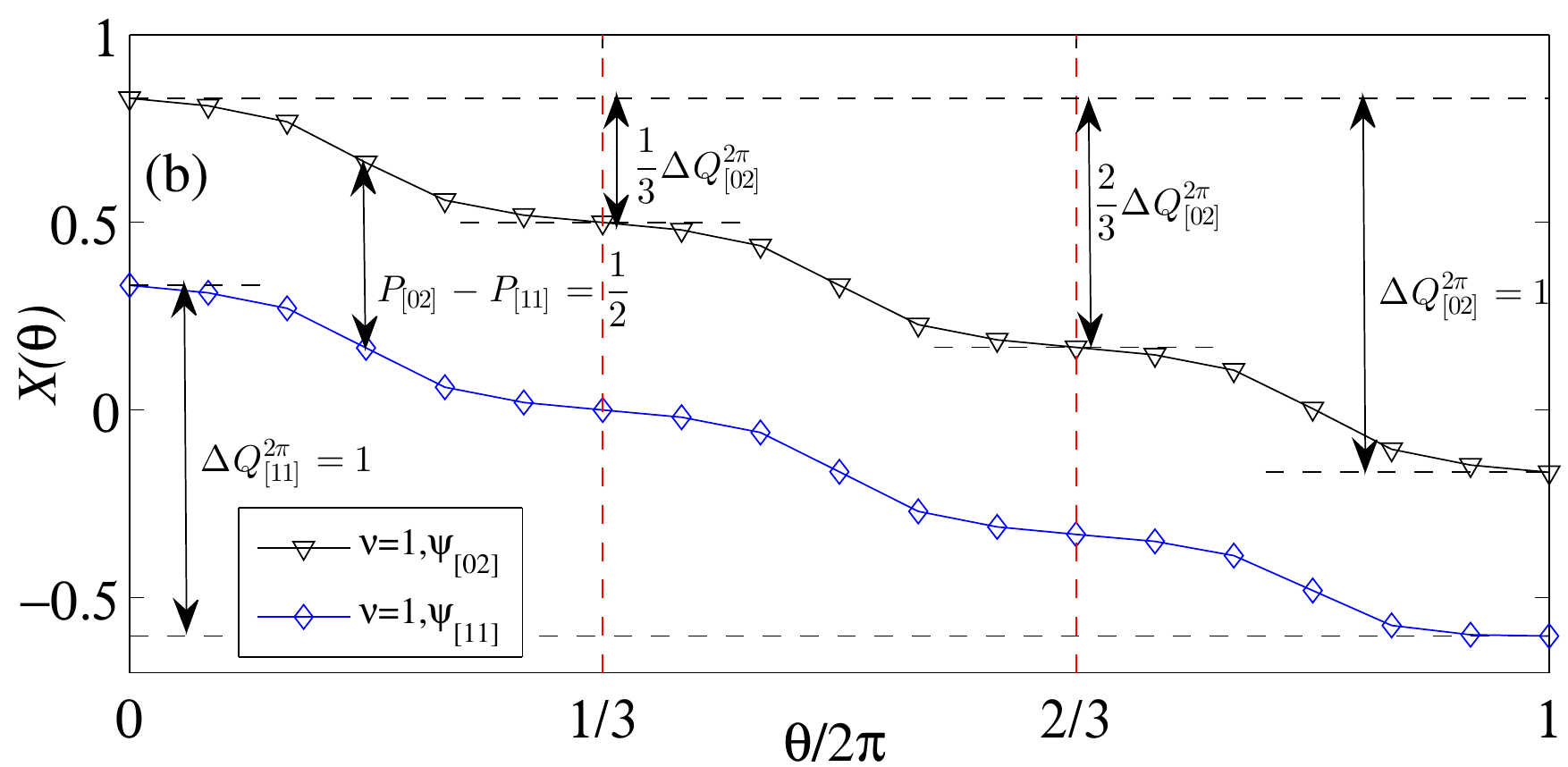}
  \includegraphics[height=1.5in,width=3.2in]{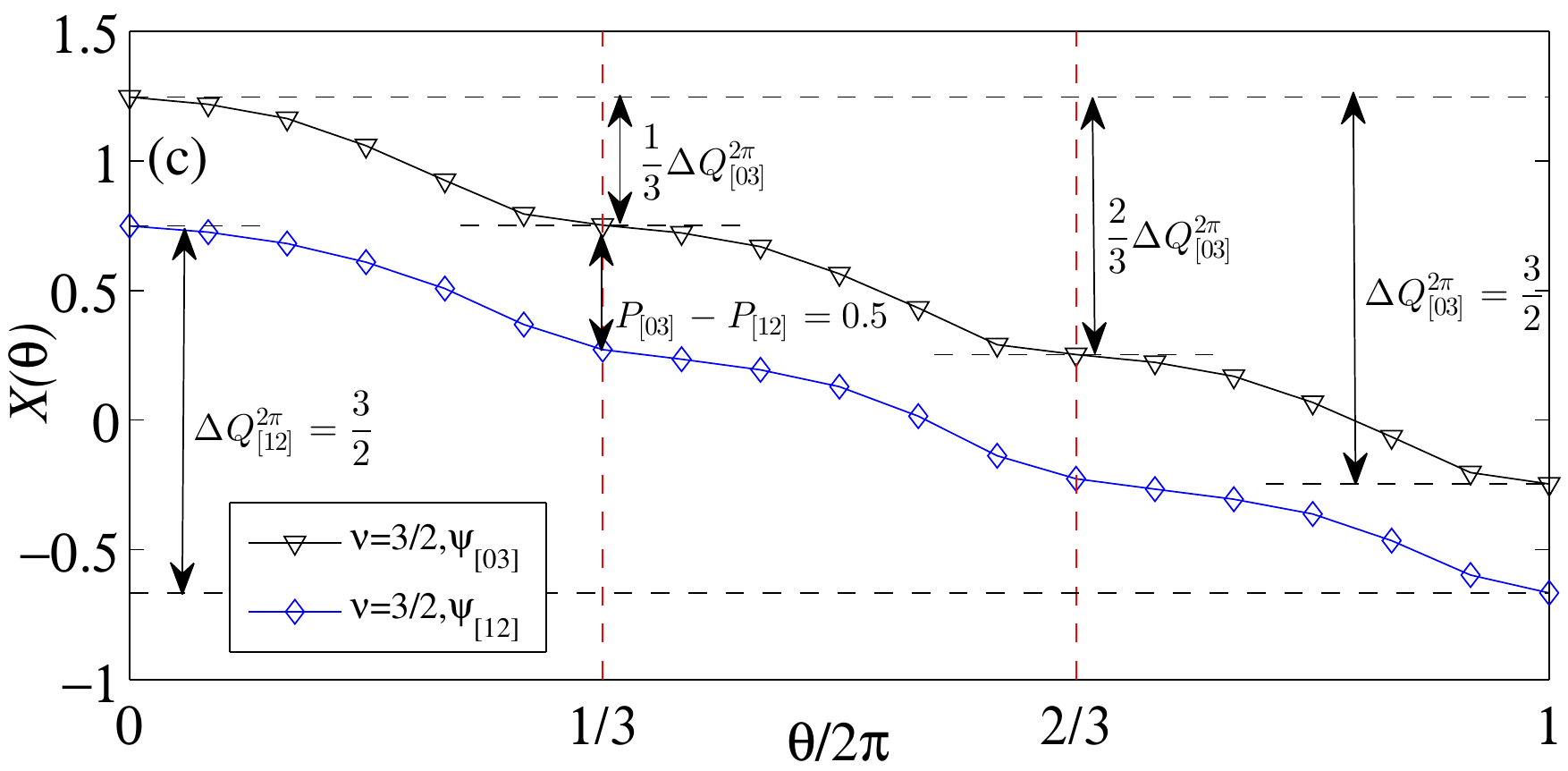}
  \caption{\label{pump}(Color online) Evolution of charge polarization in a pump cycle ($\theta=0\rightarrow 2\pi$ ) for (a) two-body hardcore bosons $H_0+\sum_{i,j}D(i-j)n_i n_j$ at $\nu=1/2,N=25,L=150,U_2=\infty,D(1)=D(2)=D(3)=2t$; (b) three-body hard-core bosons $H_0+H_3$ at $\nu=1,N=24,L=72,U_3=\infty,V_1=V_2=V_3=10t$; (c) four-body hard-core bosons $H_0+H_4$ at $\nu=3/2,N=36,L=72,U_4=\infty,W_1=W_2=W_3=10t$. The expected quantized charge pump $-\Delta Q^{2\pi}_\alpha=X_{\alpha}(2\pi)-X_{\alpha}(0)$ is also shown.}
\end{figure}

To quantify topological charge pumping and simulate the potential measurements in cold atom experiments, we introduce the charge polarization per site $X_{\alpha}$:
\begin{align}
  X_{\alpha}(\theta)=\langle\psi_{\alpha}(\theta)|\frac{1}{L}\sum_{j=1}^{L} \left(j-\frac{L+1}{2}\right) n_{j}|\psi_{\alpha}(\theta)\rangle,
\end{align}
where $|\psi_{\alpha}\rangle$ is the ground state of the root configuration sector $\alpha$.
In Figs.~\ref{pump}(a-c), we plot the evolutions of the charge polarization per site $X_{\alpha}(\theta)$ at different fillings.
Remarkably, we find the following quantized relationships governing by gauge invariance.
First of all, it is found that the drift of charge polarization $X_{\alpha}(\theta)$ in a complete pump cycle,
is always equal to the charge pumping by
\begin{align}
X_{\alpha}(2\pi)-X_{\alpha}(0) =-\Delta Q^{2\pi}_{\alpha}, \label{eq1}
\end{align}
where $\Delta Q^{2\pi}_{\alpha}$ is the charge transfer in one flux quanta.
According to the bulk-edge correspondence~\cite{QNiu1985,Hatsugai2016}: $\sigma^{\alpha}_{H}=C_{\alpha}\frac{e^2}{h}=\Delta Q^{2\pi}_{\alpha} \frac{e^2}{h}$ ($e^2/h$ is the conductance unit), we obtain the quantization of $\Delta Q^{2\pi}_{\alpha}$ is protected by the topological Chern number $C_\alpha$ of ground state. Secondly, at a given $\theta$, the difference of $X_{\alpha}(\theta)$ between two ground states are quantized to
\begin{align}
X_{\alpha}(\theta)-X_{\alpha'}(\theta)=P_{\alpha}-P_{\alpha'}, \label{eq2}
\end{align}
where $P_{\alpha}$ is defined as the intrinsic charge polarization of the topological sector $\alpha$ (see the Appendix~\ref{app3} for details), and $P_{\alpha}-P_{\alpha'}$ an intrinsic constant revealing the fractional charge of elementary excitations~\cite{Ardonne2008}. Thirdly, $X_{\alpha}(\theta)$ hosts additional hyperfine quantization structure as a function of $\theta$. That is, when $\theta$ equals to $\theta=2\pi p/q$ ($p\in Z$), the drift of charge polarization $X_{\alpha}(\theta)$ of the ground state is quantized to
\begin{align}
  X_{\alpha}(\frac{2\pi p}{q})-X_{\alpha}(0)=-\frac{p}{q}\Delta Q^{2\pi}_{\alpha}=\frac{p}{q}(X_{\alpha}(2\pi)-X_{\alpha}(0)). \label{eq3}
\end{align}
This relationship is due to the underlying $2\pi/q$-periodic Berry curvature~\cite{Marra2015},
which is an intrinsic symmetry of the Hamiltonain (Eq.~\ref{eq:ham0}) under the operation $(j,\theta+2\pi/q)\rightarrow(j+1,\theta)$ and robust even in the presence of strong interactions. In the absence of this symmetry, this symmetry-protected hyperfine fractional charge pumping would be broken.
For comparison, we calculate the charge pumping in the BEC regime, and the evolution of charge pumping in each pump cycle is nonquantized, as expected~\cite{Lu2016}. The above relationships Eqs.~(\ref{eq1}-\ref{eq3}) hold for all ground states at $\nu=k/2$ including $k=1,2,3$, as illustrated in Fig.~\ref{pump}.

In relation to the quantized behavior of the charge polarization, we investigate the quantization of the pumped charge from the entanglement spectrum (ES). ES is defined as the eigenvalues of reduced density matrix, which is obtained by partitioning the system into two halves at the chain center and tracing out the right part. The adiabatic flux insertion shifts the low levels of the ES in different charge sectors $\widehat{Q}$. The charge transfer of the total charge from the right side to the left side is encoded by $Q_{\alpha}^{\theta}=tr[\widehat{\rho}_L(\theta)\widehat{Q}]$ ($\widehat{\rho}_L$ the reduced density matrix of the left part)~\cite{Zaletel2014,Zhu2015}. In DMRG, we numerically find that for ground states at $\nu=1/2$, by threading one flux, the ES of $\psi_{[01]}$ ($\psi_{[10]}$) evolves into that of $\psi_{[10]}$ ($\psi_{[01]}$). The total charge transferred equals to $\sum_{\alpha}Q_{\alpha}^{2\pi}-Q_{\alpha}^{0}=1=\sum_{\alpha}C_{\alpha}$. For ground states at $\nu=1$, the ES of $\psi_{[
02]}$ ($\psi_{[20]}$) evolves into that of $\psi_{[20]}$ ($\psi_{[02]}$) while the ES of $\psi_{[11]}$ recovers itself after one flux quanta is threaded. We obtain the topological pumped charge $\Delta Q_{[11]}^{2\pi}=Q_{[11]}^{2\pi}-Q_{[11]}^{0}=1=C_{[11]}$, and it just equals to the change of charge polarization $X_{[11]}(2\pi)-X_{[11]}(0)$ in the main text. And the total pumped charge equals to $\sum_{\alpha}Q_{\alpha}^{2\pi}-Q_{\alpha}^{0}=3=\sum_{\alpha}C_{\alpha}$.
\section{summary and discussion}
In summary, we have shown that the charge pumping in a fractionally occupied one-dimensional lattice,
driven by adiabatically changing the periodically modulated onsite potential, is quantized to be a fractional value relating to the Chern number of the ground state.
The fractional quantization of charge pumping can be observed from a quantization of the charge polarization per site,
which can be experimentally detected using an {\it in situ} image of the center-of-mass of the gas cloud.
In current experiments with bosonic polar molecules, the dipolar interaction strength $\sim1\text{k}$ hHz when trapped in a 532nm-lattice~\cite{Lohse2016},
and the hopping amplitude $t$ can be controlled by Rabi frequency~\cite{Gerbier2010}. Thus the dominant three-body interactions can be of the same energy scale as $t$,
providing promising candidates for realizing those one-dimensional phases with nontrivial pumping.

\begin{acknowledgements}
W.Z. thanks F.~D.~M. Haldane for elucidating the calculations based on Jack polynomials, and Z. Liu, J. Wang for fruitful discussion. This work is supported by the National Science Foundation (NSF) through the grant DMR-1408560 (W.Z., D.N.S, T.S.Z.). We also acknowledge NSF grant DMR-1532249 for computational resource.
\end{acknowledgements}
\textit{Note added.}---Recently, we became aware of a similar work, Ref.~\cite{Angelakis2016}, implementing the integer quantized charge pumping of photons with attractive interactions in one-dimensional superlattice.

\appendix
\section{Quasihole Statistics}\label{app1}

\begin{figure}[t]
  \includegraphics[height=2.8in,width=3.4in]{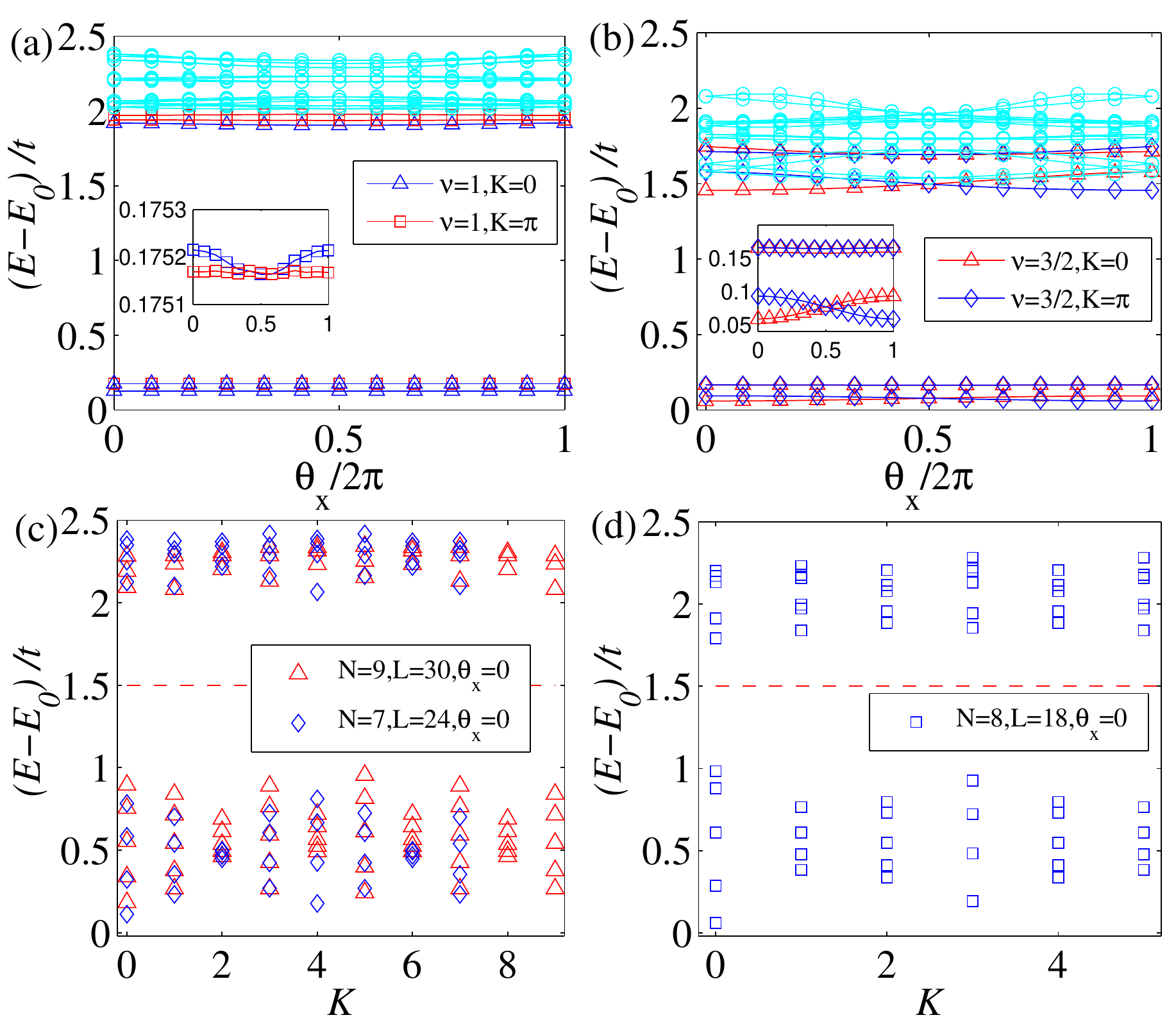}
  \caption{\label{fluxhole}(Color online) Low energy spectra flux for : (a) three-body hardcore bosons $H_0+H_3$ at $\nu=1,N=8,L=24,V_1=V_2=V_3=10t$; (b) four-body hardcore bosons $H_0+H_4$ at $\nu=3/2,N=9,L=18,W_1=W_2=W_3=10t$. Low energy spectrum for : (c) three-body hardcore bosons by removing one particle at $V_1=V_2=V_3=10t$. The number of states below the red dashed line is $N_s/2$ per momentum sector; (d) four-body hardcore bosons by removing one particle at $W_1=W_2=W_3=10t$. The total number of states below the red dashed line is 27. The parameter $\theta=0$.}
\end{figure}

In this section we describe the quasihole statistics of one dimensional systems carrying fractional charge, whose level counting also gives a close relation to its counterparts in two dimensions.  With twisted boundary condition $|\psi(x+L)\rangle=e^{i\theta_x}|\psi(x)\rangle$, the Chern number can be defined in the parameter plane $(\theta_x,\theta)$ by $C=\frac{1}{2\pi}\int_{0}^{2\pi}d\theta_x\int_{0}^{2\pi}d\theta F(\theta_x,\theta)$, where the Berry curvature $F(\theta_x,\theta)=\mathbf{Im}(\langle{\partial_{\theta_x}\psi}|{\partial_{\theta}\psi}\rangle
-\langle{\partial_{\theta}\psi}{\partial_{\theta_x}\psi}\rangle)$. For $\nu=1$ MR-like states, the low energy spectra flux vs $\theta_x$ is plotted in Fig.~\ref{fluxhole}(a). The energy of this system recover itself after one flux quanta is inserted, and the three ground states do not either mix with higher energy levels or evolve into each other. For $\nu=3/2$ RR-like states their low energy spectra flux vs $\theta_x$ is plotted in Fig.~\ref{fluxhole}(b). The energy of this system recover itself only after two flux quanta are inserted, and the four ground states do not either mix with higher energy levels.

For two-dimensional MR states, the $(2,2)$-admissible rules claim that no more than one particle is allowed to occupy within any 2 consecutive orbits. Thus, by removing one particle, the number of quasihole states of $N=N_{s}-1$ particles in $N_{s}$ orbitals reads as $N_{s}^{2}/2$, namely $N_{s}/2$ per momentum sector for $\nu=1$ bosonic MR Pfaffian state~\cite{Wang2012}. In one-dimensional systems, As shown in Fig.~\ref{fluxhole}(c), we compute the spectrum of low-energy two-quasihole state which lies in a low-energy manifold (quasihole states) separated by a gap from higher states, and find their number matches theoretical analysis. Physically, when projected into the lowest subband, the projected interaction Hamiltonian contains the two types of terms $U_3 (\beta_{k}^{\dag})^3\beta_{k}^{3}+V_3\beta_{k+1}^{\dag}(\beta_{k}^{\dag})^{2}\beta_{k}^{2}\beta_{k+1}+h.c.$, where $\beta_k$ is the particle number operator of single particle Bloch state in the lowest subband. They prevent the occupancy patterns $\cdots3\cdots,\cdots21\cdots$, therefore the admissible patterns are just the three-fold root patterns of MR-like states~\cite{Papic2014}.

For two-dimensional RR states, the $(2,3)$-admissible rules claim that no more than three particles is allowed to occupy within any 2 consecutive orbits. Thus, by removing one particle, a simple analysis of the quasihole configuration of $N=8$ particles in $N_{s}=6$ orbitals gives 5 types of configurations $203030,202121,112112,203021,211121$. Due to translational symmetry, we finally obtain 5 states for even $K$ sector, and 4 for odd $K$ sector. In one-dimensional systems, as shown in Figs.~\ref{fluxhole}(d), we compute the spectrum of low-energy quasihole state which lies in a low-energy manifold separated by a gap from higher states, and find that their number exactly matches theoretical analysis. By adding one flux quanta, we get similar results. Physically, when projected into the lowest subband, the projected interaction Hamiltonian contains the three types of terms $U_4 (\beta_{k}^{\dag})^4\beta_{k}^{4}+W_3\beta_{k+1}^{\dag}(\beta_{k}^{\dag})^{3}\beta_{k}^{3}\beta_{k+1}
+W_3(\beta_{k+1}^{\dag})^{2}(\beta_{k}^{\dag})^{2}\beta_{k}^{2}\beta_{k+1}^{2}+h.c.$. They prevent the occupancy patterns $\cdots4\cdots,\cdots31\cdots,\cdots22\cdots$, therefore the admissible patterns are just the four-fold root patterns of RR-like states~\cite{Papic2014}.

\section{Crystalline Order}\label{app2}

\begin{figure}[t]
  \centering
  \includegraphics[height=1.7in,width=3.4in]{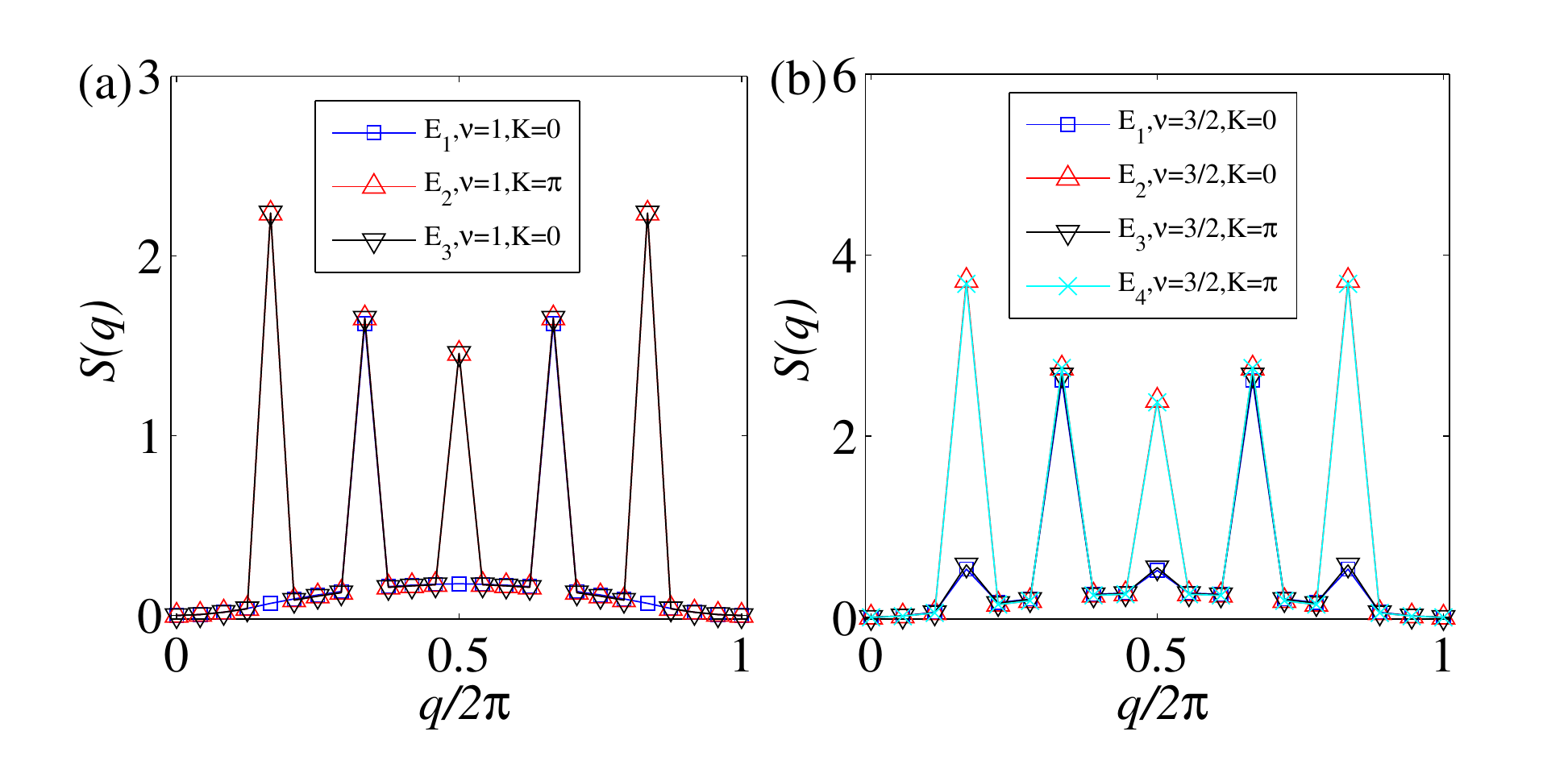}
  \caption{\label{density}(Color online) Numerical results for static density structure factors at $\theta_x=\theta=0$: (a) $S(q)$ of three gapped MR-like ground states of three-body hardcore bosons $H_0+H_3$ at $\nu=1,N=8,L=24,V_1=V_2=V_3=10t$; (b) $S(q)$ of four gapped RR-like ground states of four-body hardcore bosons $H_0+H_3$ at $\nu=3/2,N=9,L=18,W_1=W_2=W_3=10t$.}
\end{figure}

In this section we discuss the one dimensional crystalline nature of those many-body ground states in the main text. For $\nu=1$ MR-like states, however the density structure factors $S(q)=\frac{1}{L}\sum_{j,j'}e^{iq\cdot(j-j')}\left(\langle n_{j}n_{j'}\rangle-\langle n_{j}\rangle\langle n_{j'}\rangle\delta_{q,0}\right)$ of three ground states do not exhibit the same Bragg peaks. Instead, as shown in Figs.~\ref{density}(a), the structure factor of the lowest ground state only hosts $q=2\pi/3,4\pi/3$ peaks, while the other two host integer multiple times $q=2\pi/6$ vector, due to onsite double pairing occupancy. The pairing nature could be obtained from double occupancy $\langle n_{j}^{2}\rangle$ and pair correlations $\langle b_{i}^{\dag}b_{i}^{\dag}b_{j}b_{j}\rangle$.

Similarly, for $\nu=3/2$ RR-like states, as shown in Figs.~\ref{density}(b), the structure factors of four ground states do not exhibit the same Bragg peaks. Two of them (one in $(0)$ sector while the other in $(\pi)$ sector) host $q=2\pi/3,4\pi/3$ peaks, while the other two host integer multiple times $q=2\pi/6$ vector, due to onsite triple pairing occupancy which can be measured from triple occupancy $\langle n_{j}^{3}\rangle$. One distinction from Moore-Read like states is that due to inequivalent $2121\cdots$ occupancy, there also exist small bumps at $q=2\pi/6,q=\pi,q=10\pi/6$ in the structure factors of the former two.

\section{Intrinsic Charge Polarization}\label{app3}

\begin{table}[t]
\caption{\label{partition}The partition at $\nu=1$ with $N$ particles}
\begin{tabular}{|c|c|c|c|c|c|c|c|}
  \hline\hline
  $\lambda$ & $-\frac{2N-1}{2}$ & $-\frac{2N-3}{2}$ & $\cdots$ & $-\frac{7}{2}$ & $-\frac{5}{2}$ & $-\frac{3}{2}$ & $-\frac{1}{2}$
  \\\hline
  $n_{[20]}(\lambda)$ & 2 & 0 & $\cdots$ & 2 & 0 & 2 & 0 \\\hline
  $n_{[02]}(\lambda)$ & 0 & 2 & $\cdots$ & 0 & 2 & 0 & 2 \\\hline
  $n_{[11]}(\lambda)$ & 1 & 1 & $\cdots$ & 1 & 1 & 1 & 1 \\\hline
  \hline
\end{tabular}
\end{table}

\begin{table}[t]
\caption{\label{tab:P}Expected Chern number $C_{\alpha}$ and charge polarization $P_{\alpha}$ of each ground state and charge transfer $\Delta Q^{2\pi}_{\alpha}$ in a complete pump cycle $\theta=0\rightarrow 2\pi$.}
\begin{tabular}{c | c c c c }
\hline\hline
 $\nu$  &  $[\alpha]$ \kern10pt & $C_\alpha$ \kern10pt& $\Delta Q^{2\pi}_{\alpha}$ \kern10pt & $P_\alpha$ \kern10pt \\ \hline
$\nu=1/2$ & [01] &$1/2$ & $1/2$ &$1/4$ \\
          & [10] &$1/2$ & $1/2$ &$-1/4$ \\
\hline
$\nu=1$ & [02] &$1$ & $1$ &$1/2$ \\
        & [20] &$1$ & $1$ &$-1/2$ \\
        & [11] &$1$ & $1$ &$0$ \\
\hline

$\nu=3/2$ & [03] &$3/2$ & $3/2$ & $3/4$ \\
          & [30] &$3/2$ & $3/2$ & $-3/4$ \\
          & [12] &$3/2$ & $3/2$ & $1/4$ \\
          & [21] &$3/2$ & $3/2$ & $-1/4$ \\
\hline
\hline
\end{tabular}
\end{table}

We introduce here a specific quantity characterizing a given topological root configuration. For each root configuration at $\nu=1$, the partition from Jack polynomials is given by Table.~\ref{partition}. For general fillings $\nu=k/2$, the intrinsic charge polarization of the root configuration is defined as
\begin{align}
  P_{\alpha}=\frac{\nu}{N}\sum_{\lambda=-1/2}^{-N/\nu+1/2}\left[n_{\alpha}(\lambda)-\nu\right]\times\lambda. \label{charge}
\end{align}
where $n_{\alpha}(\lambda)$ is the occupation factor of the root sector $\alpha$ at position $\lambda$. In Table.~\ref{tab:P}, we give the typical values $P_{\alpha}$ from Eq.~\ref{charge}. Note that $P_{\alpha}$ is indeed a conserved invariant for any root configuration satisfying the ``sqeezing" rule $\lambda_1+\lambda_2=\lambda_1'+\lambda_2'$~\cite{Bernevig2008}. Using exact diagonalization, at $\theta=0$ we verify that $X_{\alpha}(0)-X_{\alpha'}(0)=P_{\alpha}-P_{\alpha'}$ hold for the three ground states of $\nu=1$ MR-like states; Similarly, for $\nu=3/2$ RR-like states, $X_{\alpha}(0)-X_{\alpha'}(0)=P_{\alpha}-P_{\alpha'}$ hold for the four ground states; we also obtain $X_{\alpha}(0)-X_{\alpha'}(0)=P_{\alpha}-P_{\alpha'}$ for the two ground states at $\nu=1/2$ for two-body hardcore bosons. Here, the minimal difference of the charge polarization among the ground states is just the elementary excitations's charge unit $e^{\star}=e/(kM+2)$ at fillings $\nu=k/(kM+2)$ with $(k+1)$-body interactions (Here,
we take $M=0$ for bosons)~\cite{Ardonne2008}.

\section{Effective Interaction Potential}
In this section, we consider the reduced effective interaction potentials of polar molecules in a microwave field. The internal structure of polar molecules with a closed shell electronic structure $^{1}\Sigma$ is given by the rotational degree of freedom $|J,M\rangle$. The interaction between the polar molecules at sites $\rr_i$ and $\rr_j$ is determined by the dipole-dipole interaction
\begin{align}
  V_{dd}(\rr_i-\rr_j)=\frac{d_id_j}{|\rr_i-\rr_j|^3}-\frac{3d_i\cdot(\rr_i-\rr_j)d_j\cdot(\rr_i-\rr_j)}{|\rr_i-\rr_j|^5}.\nonumber
\end{align}
Applying a circular polarized microwave field along z-axis would couple the ground state $|0,0\rangle$ with the first excited rotational level $|1,1\rangle$ by forming a dark state $|+\rangle=\alpha|0,0\rangle+\beta|1,1\rangle$. In the new representation, the effective interaction becomes
\begin{align}
  V_{dd}^{eff}=&\frac{J_{\perp}}{|\rr_i-\rr_j|^3}\big[\frac{1}{2}S_{i}^{+}S_{j}^{-}+h.c.\nonumber\\
  &+(\eta_g P_i+\eta_e Q_i)(\eta_g P_j+\eta_e Q_j)\big]\nonumber
\end{align}
where $J_{\perp}$ is the dipole coupling, $P_i=1/2+S_{i}^{z}$ and $Q_i=1/2-S_{i}^{z}$ are the projectors on the ground $|0,0\rangle$ and excited $|1,1\rangle$ states, $\eta_g,\eta_e$ induced dipole coupling coefficients. Here, we focus our interests on the lowest single-particle energy level $|+\rangle$, with on-site double occupancies and on-site Hubbard interactions taken into account. In the perturbation expansion of dipole coupling, we only keep the two-site terms (omitting three-site terms $n_in_jn_k$)
\begin{align}
  &V_{int}(\rr_i-\rr_j)=(\lambda_1\frac{J_{\perp}}{|\rr_i-\rr_j|^3}+\lambda_2\frac{J_{\perp}}{|\rr_i-\rr_j|^6})n_{+}^i n_{+}^j\nonumber\\
  &+\gamma_1\frac{J_{\perp}}{|\rr_i-\rr_j|^6}n_{+}^i n_{+}^i n_{+}^j +\gamma_2\frac{J_{\perp}}{|\rr_i-\rr_j|^9}n_{+}^i n_{+}^i n_{+}^jn_{+}^j. \nonumber
\end{align}

%

\end{document}